\documentclass[10pt]{article}
\usepackage[dvips]{color}
\usepackage{epsfig}
\usepackage{amsmath}
\usepackage{graphicx}
\usepackage{amssymb}
\usepackage{float}

\usepackage{titling}
\thanksmarkseries{alph}

\begin{document}
\title{\bf Parallel-plate and spherical capacitors in Born-Infeld electrostatics: An analytical study}
\author{S.K. Moayedi\thanks{e-mail:
s-moayedi@araku.ac.ir}\hspace{1mm} and M. Shafabakhsh\thanks{e-mail: m-shafabakhsh@phd.araku.ac.ir}\hspace{1mm}
\\
{\small {\em  Department of Physics, Faculty of Sciences,
Arak University, Arak 38156-8-8349, Iran}}\\
}
\date{\small{}}
\maketitle

\begin{abstract}
\noindent
 In 1934, Max Born and Leopold Infeld suggested and
developed a nonlinear modification of Maxwell electrodynamics, in
which the electrostatic self-energy of an electron was a finite
value. In this paper, after a brief introduction to Lagrangian
formulation of Born-Infeld electrodynamics with an external source,
the explicit forms of Gauss's law and the electrostatic energy
density in Born-Infeld theory are obtained. The capacitance and the
stored electrostatic energy for a parallel-plate and spherical
capacitors are computed in the framework of Born-Infeld
electrostatics. We show that the usual relations
$U=\frac{1}{2}C_{_{\textrm{Maxwell}}}(\triangle \phi)^{2}$ and
$U=\frac{q^{2}}{2C_{_{\textrm{Maxwell}}}}$ are not valid for a
capacitor in Born-Infeld electrostatics. Numerical estimations in
this research show that the nonlinear corrections to the capacitance
and the stored electrostatic energy for a capacitor in Born-Infeld
electrostatics are considerable when the potential difference
between the plates of a capacitor is very large.

\noindent
\hspace{0.35cm}



\end{abstract}

\section{Introduction}
As we know, Maxwell electrodynamics is a very successful theory
which is able to describe a wide range of macroscopic phenomena in
classical electricity and magnetism. Unfortunately, in Maxwell
electrodynamics the electric field of a point charge $q$ at the
position of the point charge and the classical self-energy of a
point charge are infinite, \textit{i.e.},
\begin{subequations}
\begin{eqnarray}
 \textbf E(\textbf
x)&=&\frac{q}{4\pi\epsilon_{0}|\textbf x|^{2}}\frac{\textbf
x}{|\textbf x|}\; \stackrel{|\textbf
x|\rightarrow0}{-\!\!-\!\!\!\longrightarrow} \infty,\\
U&=&\frac{q^2}{8\pi\epsilon_{0}}\int^{\infty}_{0}\frac{d|\textbf
x|}{|\textbf x|^2}\rightarrow\infty.
\end{eqnarray}
\end{subequations}

More than 80 years ago, Max Born and Leopold Infeld suggested a
nonlinear modification of Maxwell electrodynamics [1]. The
Lagrangian density of Born-Infeld electrodynamics is not only
Lorentz invariant, but also depends on the gauge potential and its
first derivative. In contrast with the Maxwell Lagrangian density,
the Born-Infeld Lagrangian density contains quartic and higher-order
powers of the electromagnetic field tensor [1-7]. In Born-Infeld
electrodynamics, the classical self-energy of a point charge like
electron is a finite value [1]. In ref. [8], the authors have
proposed a non-Abelian generalization of Born-Infeld theory from the
viewpoint of non-commutative geometry. Recent studies in string
theory show that the low-energy dynamics of a $D9$-brane, induced by
the quantum theory of the open strings attached to it, can be
described by the following action [9-13]:
\begin{equation}
S_{_{\textrm{BI}}}=\frac{1}{(4\pi^{2}\alpha^{\prime})^{5}g_{_{s}}}\;\int
d^{10}x\sqrt{-det_{_{_{_{\substack{\!\!\!\!\!\!\!\!\!\mu,\nu}}}}}\big(\eta_{\mu\nu}+T^{-1}F_{\mu\nu}\big)},
\end{equation}
where $\alpha^{\prime}$ is the ``universal Regge slope'', $g_{_{s}}$
is the ``string coupling constant'',
$T=\frac{1}{2\pi\alpha^{\prime}}$ is the ``string tension'', and
$F_{\mu\nu}$ is the ``field strength of the gauge fields living on
the $D9$-brane worldvolume'' (see ref. [13]). In ref. [13], Szabo
has shown that for a pure electric field configuration the absolute
value of the electric field in action (2) can not go beyond the
maximum value $E_{_{c}}=T=\frac{1}{2\pi\alpha^{\prime}}$. In ref.
[14], the Lagrangian formulation of Born-Infeld electrodynamics
coupled to an axionic field in the presence of an external source
has been studied. The electric field and the stored electrostatic
energy per unit length for an infinite charged line and an
infinitely long cylinder in Born-Infeld electrostatics have been
calculated analytically in ref. [15]. The black hole solutions of
Einstein's gravity in the presence of different models of nonlinear
electrodynamics in a $3+1$-dimensional space-time have been studied
in refs. [16-20]. Another interesting theory of nonlinear
electrodynamics was suggested and developed by German physicists in
1930s [21-23]. Heisenberg and his students Euler and Kockel showed
that classical electrodynamics must be corrected by nonlinear terms
due to the vacuum polarization effects [21-28]. The
Heisenberg-Euler-Kockel effective Lagrangian density is given by
\footnote{SI units are used throughout the remainder of this paper.}
[21-28]
\begin{equation}
{\cal
L}_{_{\textrm{HEK}}}=\underbrace{\frac{1}{2}\big(\epsilon_{0}\textbf
{E}^{2}-\frac{\textbf{B}^{2}}{\mu_{0}}
\big)}_{Maxwell\;term}+\underbrace{\frac{2\alpha^{2}\hbar^{3}\epsilon_{0}^{2}}{45m_{e}^{4}c^{5}}\Big[\big(\textbf{E}^{2}-c^{2}\textbf{B}^{2}\big)^{2}+7c^{2}(\textbf{E}.\textbf{B})^{2}\Big]}_{one-loop
\;corrections} +...\;,
\end{equation}
where $m_{e}$ is the electron mass, and
$\alpha=\frac{e^{2}}{4\pi\epsilon_{0}\hbar c}$  is the fine
structure constant. The general expression for effective Lagrangian
density of quantum electrodynamics can be written as follows:
\begin{equation}
{\cal L}=\sum_{i=0}^{\infty}\sum_{j=0}^{\infty}c_{_{i,j}}F^{i}G^{j},
\end{equation}
where $F:=\big(\epsilon_{0}\textbf
{E}^{2}-\frac{\textbf{B}^{2}}{\mu_{0}}\big)$ and
$G:=\sqrt{\frac{\epsilon_{0}}{\mu_{0}}}(\textbf{E}.\textbf{B})$ are
two Lorentz invariant quantities, and $c_{_{i,j}}$ are
model-dependent constant parameters [24]. It should be mentioned
that the Heisenberg-Euler-Kockel effective Lagrangian density in eq.
(3) is a very special case of eq. (4). In refs. [29,30], the authors
have suggested a new generalization of Maxwell electrodynamics which
is known as logarithmic electrodynamics. These authors have proved
that the classical self-energy of a point charge in  logarithmic
electrodynamics is a finite value. In ref. [31], the author has
suggested a new model of nonlinear electrodynamics with three
independent parameters, in which the static electric self-energy of
point-like particles is finite. In 2013, another generalization of
nonlinear electrodynamics was proposed, which is known as
exponential electrodynamics [32]. The black hole solutions of
Einstein's gravity coupled to exponential electrodynamics in a
$3+1$-dimensional space-time have been obtained in ref. [32]. In
ref. [33] and independently in ref. [34], the authors have suggested
a nonlinear generalization of Maxwell electrodynamics, in which the
electric field of a point charge is singular at the position of the
point charge but the classical self-energy of the point charge has a
finite value. In ref. [35], a novel model for nonlinear
electrodynamics has been proposed which violates the gauge symmetry.
This model describes charged particles with finite electrostatic
self-energy. Nowadays, theoretical physicists believe that the
concept of a point charge is only an idealization. The measurement
of the anomalous magnetic moment of an electron leads to the
following upper bound on the size of the electron $l$ [36]:
$l\leq10^{-17}\;\textrm{cm}.$ According to the above statements a
point charge like an electron could be considered as an extended
object with a finite size. The existence of an effective radius for
the electron leads to a finite electrostatic self-energy. In ref.
[37], the author has shown that the stored energy in a nonlinear
capacitor with a voltage-dependent capacitance does not satisfy the
usual relation $U=\frac{1}{2}C_{_{\textrm{Maxwell}}}(\triangle
\phi)^{2}$. The electric field between the plates of a
parallel-plate capacitor and the magnetic  field around a long
straight wire carrying a current $i$ have been calculated in the
framework of Heisenberg-Euler-Kockel electrodynamics [38]. Finally,
the concept of duality in Born-Infeld electrodynamics has been
studied by Gibbons and Rasheed [39]. This paper is organized as
follows. In sect. 2, the Lagrangian formulation of Born-Infeld
electrodynamics with an external source is presented. We obtain the
explicit forms of Gauss's law and the energy density of an
electrostatic field for Born-Infeld electrostatics. In sect. 3, the
capacitance and the stored electrostatic energy for a parallel-plate
capacitor are calculated nonperturbatively in the framework of
Born-Infeld theory. In sect. 4, we determine the capacitance and the
stored electrostatic energy for a spherical capacitor and an
isolated sphere in Born-Infeld electrostatics. Numerical estimations
in sect. 5 show that the nonlinear corrections to the capacitance
and the stored electrostatic energy for a capacitor in Born-Infeld
theory are negligible when the potential difference between the
plates of a capacitor is small. The metric of space-time has the
signature $(+,-,-,-)$.

\section{Lagrangian formulation of  Born-Infeld electrodynamics with an external source}
The Lagrangian density for Born-Infeld electrodynamics in a
3+1-dimensional space-time is [1]
\begin{equation}
{\cal L}_{_{\textrm{BI}}}=\epsilon_{0}\beta^{2}\bigg\lbrace
1-\sqrt{1+\frac{c^{2}}{2\beta^{2}}F_{\mu\nu}F^{\mu\nu}-\frac{c^4}{16\beta^{4}}\big(F_{\mu\nu}\,\star
F^{\mu\nu}\big)^{2}}\bigg\rbrace -J^{\mu}A_{\mu},
\end{equation}
where $F_{\mu\nu}=\partial_{\mu}A_{\nu}-\partial_{\nu}A_{\mu}$ is
the electromagnetic field tensor, $\star
F^{\mu\nu}=\frac{1}{2}\epsilon^{\mu\nu\alpha\beta}F_{\alpha\beta}$
is the dual field tensor, and $J^{\mu}=(c \rho,\textbf J)$ is an
external source for the $U(1)$ gauge field $A^{\mu}=(\frac{1}{c}
\phi , \textbf A)$ [6,39,40]. The parameter $\beta$ in eq. (5) is
called the Born-Infeld parameter. This parameter shows the upper
limit of the electric field in Born-Infeld electrodynamics. In the
limit $\beta\longrightarrow\infty $, eq. (5) reduces to the
following Lagrangian density:
\begin{equation}
{\cal L}_{_{\textrm{BI}}}|_{_{large \; \beta}}={\cal
L}_{_{\textrm{M}}}+\frac{c^{2}}{32\mu_{0}\beta^{2}}\bigg[\big(F_{\mu\nu}\,\star
F^{\mu\nu}\big)^{2}+\big(F_{\mu\nu}F^{\mu\nu}\big)^{2}\bigg]+{\cal
O}(\beta^{-4}),
\end{equation}
 where ${\cal
L}_{_{\textrm{M}}}=-\frac{1}{4\mu_{0}}F_{\mu\nu}F^{\mu\nu}-J^{\mu}A_{\mu}$
is the Maxwell Lagrangian density [40]. The Euler-Lagrange equation
for the gauge field $A_{\lambda}$ is
\begin{equation}
\frac{\partial{\cal L}_{_{\textrm{BI}}}}{\partial
A_{\lambda}}-\partial_{\rho}\bigg(\frac{\partial{\cal
L}_{_{\textrm{BI}}}}{\partial(\partial_{\rho}A_{\lambda})}\bigg)=0.
\end{equation}
If we substitute eq. (5) into eq. (7), we will obtain the
inhomogeneous Born-Infeld equations as follows:
\begin{equation}
\partial_{\rho}\bigg(\frac{F^{\rho\lambda}-\frac{c^{2}}{4\beta^{2}}\big(F_{\mu\nu} \star F^{\mu\nu}\big)\star F^{\rho\lambda}}{\sqrt{1+\frac{c^{2}}{2\beta^{2}}F_{\mu\nu}F^{\mu\nu}-\frac{c^4}{16\beta^4}\big(F_{\mu\nu} \star F^{\mu\nu}\big)^{2}}}\bigg)=\mu_{0}J^{\lambda}.
\end{equation}
Using the definition of the electromagnetic field tensor, we obtain
the following identity:
\begin{equation}
\partial_{\mu}F_{\nu\lambda}+\partial_{\nu}F_{\lambda\mu}+\partial_{\lambda}F_{\mu\nu}=0.
\end{equation}
Equation (9) is known as the Bianchi identity. In 3+1-dimensional
space-time, the components of $F_{\mu\nu}$ and $\star F^{\mu\nu}$
can be written as follows:
\begin{eqnarray}
F_{\mu\nu}&=& \left( {\begin{array}{cccc}
   0 & \frac{E_{x}}{c}\ & \frac{E_{y}}{c}  & \frac{E_{z}}{c}  \\
   -\frac{E_{x}}{c} & 0 & -B_{z} &B _{y} \\
   -\frac{E_{y}}{c} & B_{z} & 0 & -B_{x} \\
   -\frac{E_{z}}{c} & -B_{y} & B_{x} & 0
    \end{array}} \right),
\end{eqnarray}
\begin{eqnarray}
\star F^{\mu\nu}&=& \left( {\begin{array}{cccc}
   0 & -B_{x}\ & -B_{y}  & -B_{z}  \\
   B_{x} & 0 & \frac{E_{z}}{c} &-\frac{E_{y}}{c} \\
   B_{y} & -\frac{E_{z}}{c} & 0 & \frac{E_{x}}{c} \\
   B_{z} & \frac{E_{y}}{c} & -\frac{E_{x}}{c} & 0
    \end{array}} \right).
\end{eqnarray}
Using eqs. (10) and (11), eqs. (8) and (9) can be written in the
vector form as follows:
\begin{eqnarray}
\boldsymbol{\nabla} \cdot \textbf D(\textbf x ,t) &=& \rho(\textbf x
,t), \\
 \boldsymbol{\nabla} \times \textbf H(\textbf x ,t) &=& \textbf J(\textbf x ,t)+\frac{\partial
\textbf D(\textbf x ,t)}{\partial t}, \\
\boldsymbol{\nabla} \cdot \textbf B (\textbf x ,t) &=& 0, \\
 \boldsymbol{\nabla} \times \textbf E(\textbf x ,t) &=& -\frac{\partial
\textbf B(\textbf x ,t)}{\partial t},
\end{eqnarray}
where $\textbf D(\textbf x ,t)$ and $\textbf H(\textbf x ,t)$ are
given by
\begin{eqnarray}
\textbf D(\textbf x ,t) &=& \epsilon_{0}\frac {\textbf E(\textbf x ,t)+\frac{c^{2}}{\beta^{2}}\big(\textbf E(\textbf x ,t).\textbf B(\textbf x ,t)\big)\textbf B(\textbf x ,t)}{\sqrt {1-\frac{1}{\beta^{2}}\bigg(\textbf E^{2}(\textbf x ,t)-c^{2}\textbf B^{2}(\textbf x ,t)\bigg)-\frac{c^{2}}{\beta^{4}}\big(\textbf E(\textbf x ,t).\textbf B(\textbf x ,t)\big)^{2}}}, \\
 \textbf H(\textbf x ,t) &=& \frac{1}{\mu_{0}}\frac {\textbf B(\textbf x ,t)-\frac{1}{\beta^{2}}\big(\textbf E(\textbf x ,t).\textbf B(\textbf x ,t)\big)\textbf E(\textbf x ,t)}{\sqrt {1-\frac{1}{\beta^{2}}\bigg(\textbf E^{2}(\textbf x ,t)-c^{2}\textbf B^{2}(\textbf x ,t)\bigg)-\frac{c^{2}}{\beta^{4}}\big(\textbf E(\textbf x ,t).\textbf B(\textbf x ,t)\big)^{2}}}.
\end{eqnarray}

Now, let us consider the electrostatic case where $\textbf B
=\textbf J =0$ and all other quantities are independent of time. In
this case the classical Born-Infeld equations (12)-(15) are
\begin{eqnarray}
\boldsymbol{\nabla} \cdot\bigg(\frac{\textbf E(\textbf
x)}{\sqrt{1-\frac{\textbf E^{2}(\textbf x)}{\beta^{2}}}}\bigg) &=&
\frac{\rho(\textbf x )}{\epsilon_{0}},  \\
\boldsymbol{\nabla} \times \textbf E (\textbf x ) &=& 0 .
\end{eqnarray}
Equations (18) and (19) are fundamental equations of Born-Infeld
electrostatics [3,15]. Using the divergence theorem, we obtain the
integral form of eq. (18) as follows:
\begin{equation}
\oint_{S}\frac{1}{\sqrt{1-\frac{\textbf E^{2}(\textbf
x)}{\beta^{2}}}}\;\textbf E(\textbf x).\textbf{n}\; da
=\frac{1}{\epsilon_{0}} \int_{V}  \rho (\textbf x) d^{3}x ,
\end{equation}
where $V$ is the three-dimensional volume enclosed by a
two-dimensional surface $S$. Equation (20) is Gauss's law in
Born-Infeld electrostatics [15]. The symmetrized energy-momentum
tensor for Born-Infeld electrodynamics in eq. (5) is [1,41]
\begin{equation}
T^{\sigma}_{\;
\tau}=-\frac{1}{\mu_{0}\Omega}\bigg(F^{\sigma\lambda}-\frac{c^{2}}{4\beta^{2}}\big(F_{\mu\nu}
\star F^{\mu\nu}\big)\star
F^{\sigma\lambda}\bigg)F_{\tau\lambda}+\epsilon_{0}\beta^{2}(\Omega-1)\delta^{\sigma}_{\;
\tau},
\end{equation}
where $\Omega$ is defined as follows:
\begin{equation}
\Omega:=\sqrt{1+\frac{c^{2}}{2\beta^{2}}F_{\mu\nu}F^{\mu\nu}-\frac{c^{4}}{16\beta^{4}}\big(F_{\mu\nu}
\star F^{\mu\nu}\big)^{2}}.
\end{equation}
Using eqs. (10) and (11) together with eq. (21), the energy density
of an electrostatic field in Born-Infeld electrodynamics is given by
\begin{equation}
u(\textbf x )=T^{0}_{\;0}(\textbf x )=\epsilon_{0}\beta^{2}
\bigg(\frac{1}{\sqrt{1-\frac{\textbf E^{2}(\textbf
x)}{\beta^{2}}}}-1\bigg).
\end{equation}
It is necessary to note that for large values of $\beta$, the
modified electrostatic energy density in eq. (23) becomes the usual
electrostatic energy density in Maxwell electrodynamics,
\textit{i.e.},
\begin{equation}
u(\textbf x )|_{_{large \; \beta}}=\frac{1}{2}\epsilon_{0}\textbf
E^{2}(\textbf x)+{\cal O}(\beta^{-2}).
\end{equation}

\section{Capacitance of a parallel-plate capacitor in Born-Infeld electrostatics}

Let us consider two large parallel conducting plates with area $A$
and separation $d$ (see fig. 1).
\begin{figure}[ht]
\centerline{\includegraphics[width=3.2 cm]{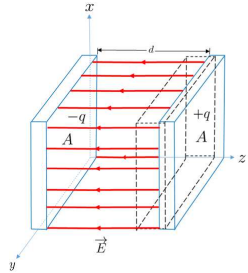}} \caption{\small A
parallel-plate capacitor. The Gaussian surface is represented by
dashed lines. $\,\hat{\textbf{e}}_{z}$ is the unit vector in the
$z$-direction.}
\end{figure}
\\
Using the modified Gauss's law in eq. (20), the electric field for
the Gaussian surface in fig. 1 becomes
\begin{eqnarray}
\textbf{E}(\textbf{x})&=& \frac{q}{\epsilon_{0}A}\frac{1}{\sqrt{1+\big(\frac{q}{\beta\epsilon_{0}A}\big)^{2}}}\,(-\hat{\textbf{e}}_{z}) \nonumber \\
\qquad &=&
-\frac{q}{\epsilon_{0}A}\,\hat{\textbf{e}}_{z}+\frac{q^{3}}{2\beta^{2}\epsilon_{0}^{3}A^{3}}\,\hat{\textbf{e}}_{z}+{\cal
O}(\beta^{-4}).
\end{eqnarray}
The first term on the right-hand side of eq. (25) shows the uniform
electric field between the plates of a parallel-plate capacitor in
Maxwell electrostatics, while the second and higher order terms in
eq. (25) show the effect of nonlinear corrections. Now, let us
calculate the potential difference $\triangle\phi$ between the
plates of a parallel-plate capacitor in Born-Infeld electrostatics.
Equation (19) implies that the electrostatic field
$\textbf{E}(\textbf{x})$ can be written as follows:
\begin{equation}
\textbf{E}(\textbf{x})=-\boldsymbol{\nabla}\phi(\textbf{x}),
\end{equation}
where $\phi(\textbf{x})$ is the electrostatic potential. Equation
(26) leads to the following formula:
\begin{equation}
\phi_{f}-\phi_{i}=-\int^{f}_{i}\textbf{E}(\textbf{x}).d\textbf{l},
\end{equation}
where $d\textbf{l}$ is an infinitesimal displacement vector. If we
use eqs. (25) and (27), we will obtain the following expression for
$\triangle\phi$:
\begin{eqnarray}
\triangle\phi&=& \phi_{+}-\phi_{-} \nonumber \\
\qquad &=&
-\int_{-}^{+}\frac{q}{\epsilon_{0}A}\frac{1}{\sqrt{1+(\frac{q}{\beta\epsilon_{0}A})^{2}}}(-\,\hat{\textbf{e}}_{z}).(\hat{\textbf{e}}_{z}\,
dz)\nonumber \\
\qquad
&=&\frac{qd}{\epsilon_{0}A}\frac{1}{\sqrt{1+(\frac{q}{\beta\epsilon_{0}A})^{2}}}.
\end{eqnarray}
We know that the amount of charge on each plate of a capacitor and
the potential difference between the plates of a capacitor are
proportional to each other, \textit{i.e.},
\begin{equation}
q=C\,\triangle\phi,
\end{equation}
where $C$ is the capacitance of the capacitor. Note that eq. (29)
has been used to determine the capacitance of nonlinear capacitors
[12,37,38]. If eq. (28) is inserted into eq. (29), the capacitance
of a parallel-plate capacitor in Born-Infeld electrostatics becomes
\begin{equation}
C_{_{\textrm{BI}}}=\frac{\epsilon_{0}A}{d}\sqrt{1+(\frac{q}{\beta\epsilon_{0}A})^{2}}.
\end{equation}
Using eqs. (28) and (30) the above expression for $C_{_{BI}}$ can be
rewritten as follows:
\begin{equation}
C_{_{\textrm{BI}}}=\frac{C_{_{\textrm{M}}}}{\sqrt{1-(\frac{\triangle\phi}{\phi_{\,\textrm{critical}}})^{2}}},
\end{equation}
where $C_{_{\textrm{M}}}=\frac{\epsilon_{0}A}{d}$ is the capacitance
of a parallel-plate capacitor in Maxwell electrostatics, and
$\phi_{\,\textrm{critical}}:= \beta d$ is the critical potential
difference. Equation (31) shows that the capacitance of a
parallel-plate capacitor in Born-Infeld electrostatics depends on
the potential difference between the plates of the capacitor. It
should be noted that the determination of $C_{_{\textrm{BI}}}$ in
eq. (31), has been mentioned as an unsolved exercise in Zwiebach's
book [12]. As we know the capacitance of a capacitor is a real
positive parameter. According to the above statement $\triangle
\phi$ in eq. (31) must satisfy the following inequality:
\begin{equation}
\triangle\phi < \phi_{\textrm{\,critical}}.
\end{equation}
The above inequality has an interesting physical interpretation. The
potential difference between the plates of a parallel-plate
capacitor in Born-Infeld electrostatics can not go beyond the
critical potential difference $\phi_{\,\textrm{critical}}$. Now, we want to
calculate the stored electrostatic energy density between the plates
of a parallel-plate capacitor in Born-Infeld electrostatics. By
putting eq. (25) in eq. (23), we obtain
\begin{equation}
u(\textbf{x})=\epsilon_{0}\beta^{2}\Big\lbrace\sqrt{1+(\frac{q}{\beta\epsilon_{0}A})^{2}}-1\Big\rbrace.
\end{equation}
Using eq. (33), the stored electrostatic energy between the plates
of a parallel-plate capacitor in Born-Infeld electrostatics
according to fig. 1 is given by

\begin{eqnarray}
U&=&\int_{{area\;of\;a\;plate}}da\;\int_{0}^{d}dz\, u(\textbf{x})\nonumber \\
\qquad
&=&\epsilon_{0}\beta^{2}\Big\lbrace\sqrt{1+(\frac{q}{\beta\epsilon_{0}A})^{2}}-1\Big\rbrace\,Ad.
\end{eqnarray}
If we use eq. (28), we can rewrite eq. (34) as follows:
\begin{eqnarray}
U&=&C_{_{\textrm{M}}}\,\phi^{2}_{\,\textrm{critical}}\bigg\lbrace \Big[1-\big(\frac{\triangle\phi}{\phi_{\,\textrm{critical}}}\big)^{2}\Big]^{-\frac{1}{2}}-1\bigg\rbrace  \nonumber \\
\qquad
&=&\underbrace{\frac{1}{2}\;C_{_{\textrm{M}}}(\triangle\phi)^{2}}_{Maxwell\;
term}+\underbrace{\frac{3}{8}\;C_{_{\textrm{M}}}\;\frac{(\triangle\phi)^{4}}{\phi^{2}_{\,\textrm{critical}}}}_{first-order\;nonlinear\;
correction}+\underbrace{{\cal
O}(\phi^{-4}_{\,\textrm{critical}})}_{higher-order\;nonlinear\;corrections}.
\end{eqnarray}
Equation (35) shows that the expression
$U=\frac{1}{2}\;C_{_{\textrm{M}}}(\triangle \phi)^{2}$ is not valid
for a parallel-plate capacitor in Born-Infeld electrostatics.

\section{Spherical capacitor in Born-Infeld electrostatics}
In this section we want to determine the capacitance of a spherical
capacitor in Born-Infeld electrostatics. Let us consider two
concentric spherical conductors of radii $a$ and $b$ (see fig. 2).
\begin{figure}[ht]
\centerline{\includegraphics[width=3.2 cm]{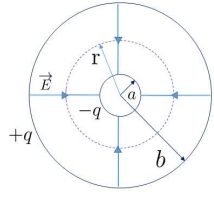}}\caption{\small
The Gaussian surface for a spherical capacitor. The spherical
symmetry of the problem implies that $\textbf{E}(\textbf{x})=
E_{r}(r)\,\hat{\textbf{e}}_{r}$, where $\hat{\textbf{e}}_{r}$ is the
radial unit vector in spherical coordinates $(r,\theta,\varphi)$.}
\end{figure}
\\
Using the spherical symmetry of the problem together with the
modified Gauss's law in eq. (20), the electric field between the
plates of a spherical capacitor becomes
\begin{equation}
\textbf{E}(\textbf{x})=\frac{q}{4\pi\epsilon_{0}r^{2}}\frac{1}{\sqrt{1+(\frac{q}{4\pi\epsilon_{0}\beta
r^{2}})^{2}}}(-\,\hat{\textbf{e}}_{r})\;;a<r<b.
\end{equation}
Using eqs. (27) and (36), the potential difference between the
plates of a spherical capacitor in Born-Infeld electrostatics can be
determined as follows:
\begin{eqnarray}
\triangle\phi&=& \phi_{+}-\phi_{-} \nonumber \\
\qquad &=& \int
_{a}^{b}\frac{q}{4\pi\epsilon_{0}
r^{2}}\frac{1}{\sqrt{1+(\frac{q}{4\pi\epsilon_{0}\beta
r^{2}})^{2}}}dr \nonumber \\
\qquad
&=&\frac{q}{4\pi\epsilon_{0}}\Big\lbrace(\frac{1}{a}-\frac{1}{b})-\frac{1}{10}s^{4}(\frac{1}{a^{5}}-\frac{1}{b^{5}})+\frac{1}{24}s^{8}(\frac{1}{a^{9}}-\frac{1}{b^{9}})+{\cal
O} (s^{12})\Big\rbrace,
\end{eqnarray}

where $s:=(\frac{q}{4\pi\epsilon_{0}\beta})^{\frac{1}{2}}$ is a
constant positive parameter with dimension $length$. After inserting
eq. (37) into eq. (29), we obtain
\begin{eqnarray}
C_{_{\textrm{BI}}} &=& 4\pi\epsilon_{0}\, \frac{ab}{b-a} \bigg\lbrace1+\frac{1}{10}s^{4}\frac{b^{5}-a^{5}}{a^{4}b^{4}(b-a)}+\frac{1}{4}s^{8}\frac{1}{a^{8}b^{8}(b-a)} \nonumber \\
\qquad & &
\Big[\frac{1}{25}\frac{(b^{5}-a^{5})^2}{(b-a)}-\frac{1}{6}(b^{9}-a^{9})\Big]+{\cal
O}(s^{12})\bigg\rbrace.
\end{eqnarray}
Substitution of eq. (36) into eq. (23) yields the following
expression for the electrostatic energy density between the plates
of a spherical capacitor
\begin{equation}
u(\textbf{x})=\epsilon_{0}\beta^{2} \bigg\lbrace
\sqrt{1+\big(\frac{s}{r}\big)^{4}}-1\bigg\rbrace.
\end{equation}
Using the above equation, the total stored energy in spherical
capacitor is given by

\begin{eqnarray}
U &=& 4\pi\epsilon_{0}\beta^{2}\,\int_{a}^{b}r^{2}\bigg\lbrace \sqrt{1+\big(\frac{s}{r}\big)^{4}}-1\bigg\rbrace dr \nonumber \\
\qquad &=&\frac{q^{2}}{8\pi\epsilon_{0}}\frac{b-a}{ab }\bigg\lbrace
1-\frac{1}{20}s^{4}\frac{b^{5}-a^{5}}{a^{4}b^{4}(b-a)}
+\frac{1}{72}s^{8}\frac{b^{9}-a^{9}}{a^{8}b^{8}(b-a)}+{\cal
O}(s^{12})\bigg\rbrace.
\end{eqnarray}

If we use eqs. (37), (38), and (40), the potential, capacitance, and
total energy for an isolated spherical conductor of radius $R$ can
be determined as follows:
\begin{equation}
\lim_{b \rightarrow \infty}\;\lim_{a\rightarrow R} \; \triangle\phi
=\frac{q}{4\pi\epsilon_{0}R}\bigg\lbrace1-\frac{1}{10}(\frac{s}{R})^{4}+\frac{1}{24}(\frac{s}{R})^{8}+{\cal
O}\Big((\frac{s}{R})^{12}\Big)\bigg\rbrace,
\end{equation}

\begin{equation}
\lim_{b \rightarrow \infty}\;\lim_{a\rightarrow R} \;
C_{_{\textrm{BI}}} = 4\pi\epsilon_{0}R
\bigg\lbrace1+\frac{1}{10}(\frac{s}{R})^{4}-\frac{19}{600}(\frac{s}{R})^{8}+{\cal
O}\Big((\frac{s}{R})^{12}\Big)\bigg\rbrace,
\end{equation}

\begin{equation}
\lim_{b \rightarrow \infty}\;\lim_{a\rightarrow R}  \;U =
\frac{q^{2}}{8\pi\epsilon_{0}R}\bigg\lbrace1-\frac{1}{20}(\frac{s}{R})^{4}+\frac{1}{72}(\frac{s}{R})^{8}+{\cal
O}\Big((\frac{s}{R})^{12}\Big)\bigg\rbrace.
\end{equation}

Equations (38) and (42) show that the capacitance of a spherical
capacitor in Born-Infeld electrostatics depends on the amount of
charge on each plate of the capacitor.

\section{Summary and conclusions}
Today, we know that the classical self-energy of a point charge has
an infinite value in Maxwell electrodynamics. In 1934, Max Born and
Leopold Infeld introduced a nonlinear generalization of Maxwell
electrodynamics, in which the classical self-energy of an electron
was a finite value [1]. In Born-Infeld electrodynamics the absolute
value of the electric field can not go beyond the critical electric
field $\beta$, \textit{i.e.}, $|\textbf E|\leq\beta$. Born and
Infeld attempted to calculate $\beta$ by equating the classical
self-energy of the electron in their theory with its rest mass
energy. They obtained the following numerical value for $\beta$ [1]:
\begin{equation}
\beta_{_{\textrm{Born-Infeld}}}=1.2\times10^{20}\frac{\textrm{V}}{\textrm{m}}.
\end{equation}
In 1973, Gerhard Soff and his coworkers obtained a new lower bound
on $\beta$ [42]. This lower bound on $\beta$ was
\begin{equation}
\beta_{_{\textrm{Soff}}}\geq
1.7\times10^{22}\frac{\textrm{V}}{\textrm{m}}.
\end{equation}
In a recent paper about photonic processes in Born-Infeld theory,
Davila \textit{et al.} [43] have obtained the following lower bound
on $\beta$:
\begin{equation}
\beta_{_{\textrm{Davila}}}\geq
2.0\times10^{19}\frac{\textrm{V}}{\textrm{m}}.
\end{equation}
It is interesting to note that the lower bound in eq. (46) is near
to the $\beta_{_{\textrm{Born-Infeld}}}$ in eq. (44). In this paper,
after a brief introduction to the Lagrangian formulation of
Born-Infeld electrodynamics in the presence of an external source,
the capacitance of parallel-plate and spherical capacitors have been
calculated analytically in the framework of Born-Infeld
electrostatics. According to eqs. (30), (31), (38), and (42) the
capacitance of a capacitor in Born-Infeld electrostatics depends on
the amount of charge on each plate of the capacitor. Using the
relation $C_{_{\textrm{M}}}=\frac{\epsilon_{0}A}{d}$, eq. (34) can
be rewritten as follows:
\begin{eqnarray}
U &=& \beta^{2}d^{2}C_{_{\textrm{M}}}\Big\lbrace\big[1+\big(\frac{q}{\beta d C_{_{\textrm{M}}}}\big)^{2}\big]^{\frac{1}{2}}-1 \Big\rbrace\nonumber \\
\qquad &=&\frac{q^{2}}{2
C_{_{\textrm{M}}}}-\frac{q^{4}}{8\beta^{2}d^{2}C_{_{\textrm{M}}}^{3}}+{\cal
O}(\beta^{-4}).
\end{eqnarray}

Equations (35), (40), (43), and (47) show that the stored energy in
a Born-Infeld capacitor does not satisfy the relations
$U=\frac{1}{2}C_{_{\textrm{M}}}(\triangle \phi)^{2}$ and
$U=\frac{q^{2}}{2 C_{_{\textrm{M}}}}$. In order to obtain a better
understanding of nonlinear effects in a parallel-plate capacitor,
let us estimate the numerical value of the following expression (see
eq. (31)):
\begin{equation}
\frac{\triangle
C^{^{^{\textrm{parallel-plate}}}}}{C^{^{^{\textrm{parallel-plate}}}}_{_{\textrm{M}}}}=\Big[1-\big(\frac{\triangle
\phi}{\phi_{\textrm{critical}}}\big)^{2}\Big]^{-\frac{1}{2}}-1,
\end{equation}
where
\begin{equation}
\triangle
C^{^{^{\textrm{parallel-plate}}}}:=C^{^{^{\textrm{parallel-plate}}}}_{_{\textrm{BI}}}-C^{^{^{\textrm{parallel-plate}}}}_{_{\textrm{M}}}.
\end{equation}
Let us assume the following approximate but realistic values (see
page 736 in ref. [44]):
\begin{equation}
A=115 \;\textrm{cm}^{2},\; d=1.24 \;\textrm{cm},\; \triangle \phi =
85.5 \;\textrm{V}.
\end{equation}
By putting eqs. (44), (45), (46), and (50) into eq. (48), we get
\begin{subequations}
\begin{eqnarray}
\triangle C^{^{^{\textrm{parallel-plate}}}}_{_{\textrm{Born-Infeld}}}\approx& 1.65\times10^{-33}C^{^{^{\textrm{parallel-plate}}}}_{_{\textrm{M}}}\approx 1.35\times 10^{-44}\;\textrm{F},\\
\triangle C^{^{^{\textrm{parallel-plate}}}}_{_{\textrm{Soff}}}\approx& 8.22\times10^{-38}C^{^{^{\textrm{parallel-plate}}}}_{_{\textrm{M}}}\approx  6.75\times 10^{-49}\;\textrm{F},\\
\triangle C^{^{^{\textrm{parallel-plate}}}}_{_{Davila}}\approx&
5.95\times10^{-32}C^{^{^{\textrm{parallel-plate}}}}_{_{\textrm{M}}}\approx
4.87\times 10^{-43}\;\textrm{F}.
\end{eqnarray}
\end{subequations}
Note that in eqs. (51b) and (51c) the minimum value of $\beta$ in
eqs. (45) and (46) has been used. Equations (51a), (51b), and (51c)
show that the nonlinear corrections to the capacitance of a
parallel-plate capacitor are negligible when the potential
difference between the plates of a parallel-plate capacitor is
small. As another example, let us estimate the numerical value of
the second term on the right-hand side of eq. (43). For this
purpose, we rewrite eq. (43) as follows:
\begin{equation}
U_{_{\textrm{BI}}}^{^{^{\textrm{isolated\;sphere}}}}=U_{_{\textrm{M}}}^{^{^{\textrm{isolated\;sphere}}}}+\triangle
U^{^{^{\textrm{isolated\;sphere}}}}+ {\cal
O}\Big((\frac{s}{R})^{8}\Big),
\end{equation}
where
\begin{equation}
U_{_{\textrm{M}}}^{^{^{\textrm{isolated\;sphere}}}}:=\frac{q^2}{8\pi\epsilon_{0}R},
\end{equation}
\begin{equation}
\triangle
U^{^{^{\textrm{isolated\;sphere}}}}:=-\frac{1}{20}\Big(\frac{s}{R}\Big)^{4}U_{_{\textrm{M}}}^{^{^{\textrm{isolated\;sphere}}}}.
\end{equation}
Using eqs. (53) and (54), the ratio of $\triangle
U^{^{^{\textrm{isolated\;sphere}}}}$ to
$U_{_{\textrm{M}}}^{^{^{\textrm{isolated\;sphere}}}}$ is given by
\begin{equation}
\Big|\frac{\triangle
U^{^{^{\textrm{isolated\;sphere}}}}}{U_{_{\textrm{M}}}^{^{^{\textrm{isolated\;sphere}}}}}\Big|=\frac{1}{20}\Big(\frac{s}{R}\Big)^{4}.
\end{equation}
Then, assuming the following approximate but realistic values for
an isolated sphere (see page 730 in ref. [44]):
\begin{equation}
R=6.85 \;\textrm{cm},\;q=1.25\times 10^{-9}\;\textrm{C}.
\end{equation}
If we put eqs. (44), (45), (46), and (56) into eq. (55), we will
obtain the following results:
\begin{subequations}
\begin{eqnarray}
\big|\triangle
U_{_{\textrm{Born-Infeld}}}^{^{^{\textrm{isolated\;sphere}}}}\big|\approx&1.74\times
10^{-73}U_{_{\textrm{M}}}^{^{^{\textrm{isolated\;sphere}}}}\approx1.79\times10^{-80}\;\textrm{J},\\
\big|\triangle
U_{_{\textrm{Soff}}}^{^{^{\textrm{isolated\;sphere}}}}\big|\approx&4.33\times
10^{-82}U_{_{\textrm{M}}}^{^{^{\textrm{isolated\;sphere}}}}\approx4.46\times10^{-89}\;\textrm{J},\\
\big|\triangle
U_{_{\textrm{Davila}}}^{^{^{\textrm{isolated\;sphere}}}}\big|\approx&2.26\times
10^{-70}U_{_{\textrm{M}}}^{^{^{\textrm{isolated\;sphere}}}}\approx2.33\times10^{-77}\;\textrm{J}.
\end{eqnarray}
\end{subequations}

As the previous numerical example, the minimum value of $\beta$ in
eqs. (45) and (46) has been used in eqs. (57b) and (57c). In fact,
as is clear from eqs. (57a), (57b), and (57c), the nonlinear
corrections to stored energy in an isolated sphere are very small
for weak electric fields. In our future works, we hope to study the
problems discussed in this research from the viewpoint of
Heisenberg-Euler-Kockel electrostatics [21-23].



\end{document}